\documentclass[prb,preprint,superscriptaddress,showpacs,preprintnumbers,amsmath,amssymb]{revtex4}

\usepackage{graphicx}
\usepackage{dcolumn}
\usepackage{bm}

\begin{document}

\preprint{APS/123-QED}

\title{Crystalline electric field effects in Ce 3$d$ core-level spectra of heavy-fermion systems: Hard X-ray photoemission spectroscopy on CeNi$_{1-x}$Co$_x$Ge$_2$}

\author{H. J. Im}
\email{hojun@cc.hirosaki-u.ac.jp}
\affiliation{Department of Physics, Sungkyunkwan University, Suwon 440-746, Korea.}
\affiliation{UVSOR Facility, Institute for Molecular Science, Okazaki 444-8585, Japan}
\affiliation{Department of Advanced Physics, Hirosaki University, Hirosaki 036-8561, Japan}

\author{T. Ito}
\affiliation{UVSOR Facility, Institute for Molecular Science, Okazaki 444-8585, Japan}
\affiliation{School of Physical Sciences, The Graduate University for Advanced Studies, Okazaki 444-8585, Japan.}

\author{S. Kimura}
\affiliation{UVSOR Facility, Institute for Molecular Science, Okazaki 444-8585, Japan}
\affiliation{School of Physical Sciences, The Graduate University for Advanced Studies, Okazaki 444-8585, Japan.}

\author{H.-D. Kim}
\affiliation{Pohang Accelerator Laboratory, Pohang University of Science and Technology, Pohang 790-784, Korea.}

\author{J. B. Hong}
\author{Y. S. Kwon}
\email{yskwon@skku.ac.kr}
\affiliation{Department of Physics, Sungkyunkwan University, Suwon 440-746, Korea.}

\date{\today}

\begin{abstract}
High-resolution hard X-ray photoemission measurements have been performed to clarify the electronic structure originating from the strong correlation between electrons in bulk Ce 3$d$ core-level spectra of CeNi$_{1-x}$Co$_x$Ge$_2$ (0 $\leq$ $x$ $\leq$ 1). In the Ce 3$d_{5/2}$ core-level spectra, the variation of satellite structures ($f^2$ peaks) shows that the hybridization strength between Ce 4$f$- and conduction electrons gradually increases with Co concentration in good agreement with the results of Ce 3$d$-4$f$ and 4$d$-4$f$ resonant photoemission spectroscopies. Particularly, in Ce 3$d_{3/2}$ core-level spectra, the multiplet structures of $f^1$ peaks systematically change with the degeneracy of $f$-states which originates from crystalline electric field effects.
\end{abstract}

\maketitle

\section{Introduction}

Hard X-ray photoemission spectroscopy (HXPES) enables us to directly observe core-level states with a large probing depth and is considered as the one of the most powerful tools to study the bulk electronic structure of strongly correlated electrons systems (SCES) \cite{Hufn95}.
Core-level spectra in a solid are results of many-body effects derived from the interaction of the core-hole state with valence and conduction bands \cite{Saka88}.
Particularly, strong correlation between electrons gives rise to the change of core-level line shapes such as satellite and multiplet structures through the exchange interaction, spin-orbit interaction, and so on \cite{Hufn95}.
There are several excellent reports on variation of satellite structures, which have been facilitated the understanding of strong correlation effects.
For example, (i) in heavy-fermion systems the satellite peaks of Ce 3$d$ core-level show the systematic change as a function of the hybridization between Ce 4$f$- and conduction electrons \cite{Gunn83}, (ii) in ruthernates which show the Mott transition, screened peaks grow as the system become more metallic \cite{Kim04}, and (iii) in high-T$_c$ cuprate superconductors the hole (electron) doping causes the well-screened features in Cu 2$p$ core-level spectra \cite{Tagu05}.
On the other hand, multiplet structures have been rarely studied due to the difficulty of precise analysis of their small spectral variation.
To investigate correlation effects in multiplet structure, the systematic studies of crystalline electric field (CEF) effects in heavy-fermion systems are very suitable.
CEF effects have been usually considered as a result from the localized character of electrons and play an important role to determine the degeneracy and orbital symmetry of $f$-states in the ground state.
This eventually influences on multiplet structures and can be experimentally observable.
Moreover, CEF effects are fundamental phenomena to decide physical properties in heavy fermion Ce compounds \cite{Cox99}. For instance, they are the one of reasons to change the hybridization strength between Ce 4$f$- and conduction electrons \cite{Taki02} and sometimes gives rise to the competition between spin and orbital fluctuation \cite{Curr01}.
In this paper, we firstly report the systematic variation of multiplet structure in Ce 3$d$ core-level spectra due to CEF effects in CeNi$_{1-x}$Co$_x$Ge$_2$ (0 $\leq$ $x$ $\leq$ 1) system, using HXPES with the high photon-energy (\textit{hv} = 7941.5 eV) and high energy-resolution ($\Delta$E = 180 meV).

\section{Samples and Experiments}

The ground state of CeNi$_{1-x}$Co$_x$Ge$_2$ changes from antiferromagnetic ($x$ = 0) to non-magnetic ($x$ = 1) regime via quantum critical point ($x$ = 0.3) in the Doniach phase diagram \cite{Lee05,Doni77}.
In particular, the degeneracy of $f$-state (N$_f$) in the ground state and Kondo temperature ($\rm T_{K}$) systematically varies due to CEF effects:
As Co concentration increases, $T_{K}$ increases from 5 to 284 K with the dramatic changes between $x$ = 0.6 ($T_{K}$ = 21 K) and 0.7 ($T_{K}$ = 110 K) and between $x$ = 0.8 ($T_{K}$ = 160 K) and 0.9 ($T_{K}$ = 200 K), where the N$_f$ changes from 2 to 4 and from 4 to 6, respectively \cite{Lee05,Im06}.

The used samples were prepared by arc-melting and annealing at 900 $^{\circ}$C for about one week.
X-ray diffraction patterns have confirmed that they are well crystallized in the orthorhombic CeNiSi$_2$-type (\textit{Cmcm}) structure and single phase.

HXPES measurements were performed at the beamline BL29XU of the SPring-8.
The clean surface was obtained by \textit{in situ} fracturing at the measurement temperature, T = 20 K.
The vacuum during measurements was below 5 $\times$ 10$^{-8}$ Pa.
The core-level spectrum of Au 4$f$-state was used to calibrate the Fermi-level.
During measurement, sample contamination was checked by monitoring the O 1$s$ peak.

\section{Results and discussion}

\begin{figure}
\includegraphics[width=85mm,clip]{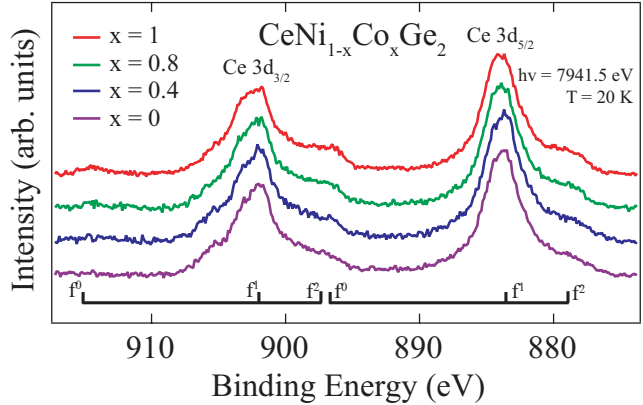}
\caption{\label{fig:figure1} (color online) Ce 3$d$ core level photoemission spectra of CeNi$_{1-x}$Co$_x$Ge$_2$ ($x$ = 0, 0.4, 0.8, 1) measured at 20 K using $hv$ = 7941.5 eV.}
\end{figure}

Figure 1 shows the bulk Ce 3$d$ core level spectra of CeNi$_{1-x}$Co$_x$Ge$_2$ ($x$ = 0, 0.4, 0.8, 1) with the probing depth of $\sim$ 200 \textrm{\AA} due to the high kinetic energy of the emitted photoelectrons ($E_k$ $\sim$ 7050 eV) \cite{Tanu87}.
Backgrounds of all spectra are subtracted by the Shirley correction \cite{Shir72} and are normalized to the intensity of the Ce 3$d_{5/2}$ $f^1$ peak.
Ce 3$d_{3/2}$ ($E_B$ $\sim$ 902 eV, where $E_B$ is the binding energy) and Ce 3$d_{5/2}$ ($E_B$ $\sim$ 883 eV) peaks are separated by about 19 eV due to the spin-orbit interaction of Ce 3$d$ state.
Each Ce 3$d$ peak consists of one main peak and two satellite peaks, which are assigned by $f^1$ and $f^0$, $f^2$, respectively, according to the large contribution to the weight of the peak \cite{Gunn83}.
Both Ce 3$d_{3/2}$ and Ce 3$d_{5/2}$ core-level spectra reveal that the $f^2$ and $f^0$ peaks gradually increase with increasing Co concentration, namely, increasing the hybridization between Ce 4$f$- and conduction electrons.
This behavior has been well explained in the frame of the single impurity Anderson model (SIAM) as confirmed in many heavy fermion Ce compounds \cite{Gunn83,Fugg83,Brai97}.
Especially, the intensity of the $f^2$ peak can be qualitatively regarded as the hybridization strength:
The fraction of initial $f^2$-states increases with increasing hybridization, and then it directly causes the $f^2$ configuration to increase in final state through the photoemission processes \cite{Gunn83}.
It is widely accepted that the hybridization strength is given by the intensity ratio , I$_{2}$ = I($f^2$)/(I($f^1$)+I($f^2$)), where I($f^{1(2)}$) is the intensity of $f^{1(2)}$ peak;
note that I($f^1$) is usually very small compared to I($f^1$) and I($f^2$) and is negligible in analysis \cite{Fugg83}.
Beyond this, in both Ce 3$d_{3/2}$ and Ce 3$d_{5/2}$ core-level spectra, we observe the fine structures of $f^1$ peak, which show the systematic variation as a function of Co concentration;
the former shows the multi-peak structure and latter the asymmetry shape.
It has been reported that they are attributed to the multiplet structure in final state \cite{Imada89,Mats08}.
In order to understand the CEF effects in multiplet structures, these spectra are analyzed by two-steps.
Firstly, we estimate the hybridization strength using Ce 3$d_{5/2}$ core level spectra because the fine structure of Ce 3$d_{5/2}$ $f^1$ peak is just asymmetric and is not prominent in comparison with those of Ce 3$d_{3/2}$ core level spectra.
Secondly, the systematic studies of multiplet structures are carried out using Ce 3$d_{3/2}$ core level spectra where there seem to be two peaks separated by the energy of about 1 eV in the $f^1$ peak.

\begin{figure}
\includegraphics[width=75mm,clip]{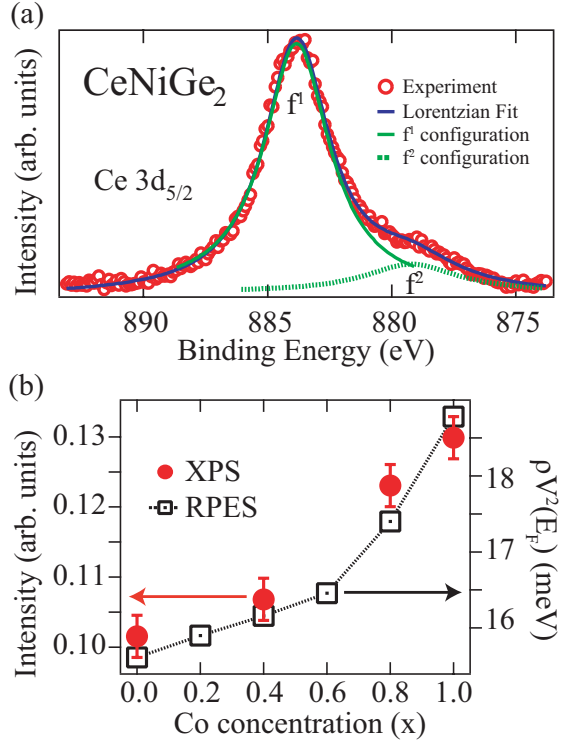}
\caption{\label{fig:figure2} (color online)
(a) Ce 3$d_{5/2}$ core level photoemission spectra of CeNiGe$_2$. The spectrum is well deconvoluted into the $f^1$ and $f^2$ configuration by Lorentzian fitting. (b) Comparison of I$_{2}$ = I($f^2$)/(I($f^1$)+I($f^2$)), where I($f^{1(2)}$) is the intensity of $f^{1(2)}$ peak, and the bulk hybridization strength obtained from the valence band resonant photoemission. ($\rho$V$^2$, where $\rho$ is the density of states of conduction band and V hybridization between Ce 4$f$- and conduction electrons)}
\end{figure}

Figure 2(a) shows the Ce 3$d_{5/2}$ core-level spectrum of CeNiGe$_2$. The spectrum is well deconvoluted by the two Lorentzian functions, which correspond to Ce 3$d_{5/2}$ $f^1$ and Ce 3$d_{5/2}$ $f^2$ peaks.
Concerning the Ce 3$d_{5/2}$ $f^0$ peak around 897 eV, we can not well distinguish it from the Ce 3$d_{3/2}$ $f^2$ peak due to the superposition (Fig. 1).
However, the influence of $f^0$ peak on analysis is negligible as mentioned above \cite{Fugg83}.
I$_{2}$ = I($f^2$)/(I($f^1$)+I($f^2$)) is estimated by the height of the each fitted Lorentzian function.
The results for all sample are plotted in Fig. 2(b), together with the bulk hybridization strength obtained from the Ce 3$d$-4$f$ and 4$d$-4$f$ resonant photoemission (RPES) of the valence band \cite{Im05}.
We recognize that the I$_{2}$ continuously increases with Co concentration through QCP ($x$ = 0.3) and steeply change from 0.4 to 1 via 0.8 in good agreement with the RPES results.
This confirms again that I$_2$ is proportional to hybridization strength and indicates that the Ce 3$d$ core-level spectra, used in this paper, are very reliable.

\begin{figure}
\includegraphics[width=80mm,clip]{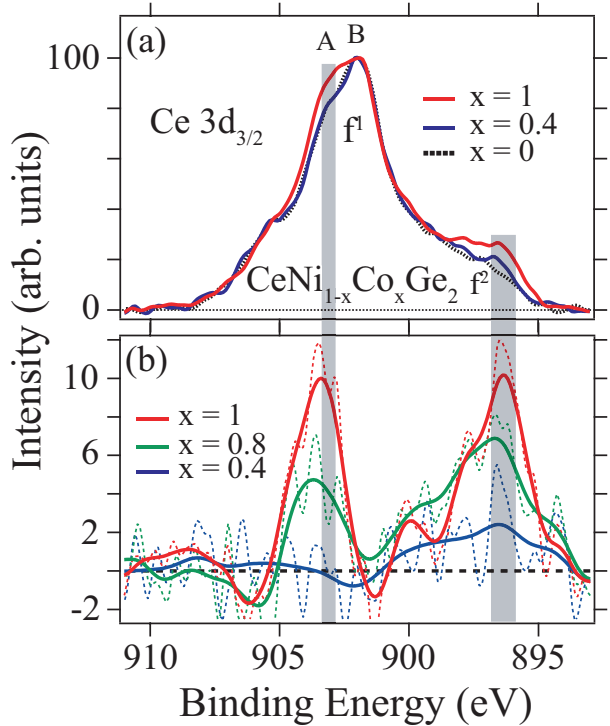}
\caption{\label{fig:figure3} (color online)
(a) The Ce 3$d_{3/2}$ core level photoemission spectra of CeNi$_{1-x}$Co$_x$Ge$_2$ (x = 0, 0.4, 1) without offset. Fine structures of $f^1$ peak are labeled by A and B. (b) Differences between the spectrum of CeNiGe$_2$ ($x$ = 0) and those of the other samples ($x$ = 0.4, 0.8, 1). Thick solid lines are obtained from the smoothing of the thin dashed line.}
\end{figure}

Figure 3(a) displays Ce 3$d_{3/2}$ core-level spectra of CeNi$_{1-x}$Co$_x$Ge$_2$ ($x$ = 0, 0.4, 1) without offset to easily compare each other.
The multiplet peak of higher binding energy in Ce 3$d_{3/2}$ $f^1$ configuration is assigned as A ($E_B$ $\sim$ 903 eV) and the lower one as B ($E_B$ $\sim$ 902 eV).
All spectra are normalized to the intensity of the B peak.
We clearly observe that the spectral weights of both A and Ce 3$d_{3/2}$ $f^2$ peaks vary as a function of Co concentration but display the different behavior.
In order to clarify the variation of A and Ce 3$d_{3/2}$ $f^2$ peaks, the spectra for $x$ = 0.4, 0.8, and 1, are subtracted from that of CeNiGe$_2$ ($x$ = 0) as shown in Fig. 3(b).
We unambiguously observe that the intensity of A peak, I(A), is the almost same between $x$ = 0 and 0.4, and suddenly increases between 0.4 and 0.8, where the top of the asymmetric Ce 3$d_{3/2}$ $f^1$ peak abruptly changes from the lower to higher binding energy as shown in Fig. 1.
And then I(A) increases from $x$ = 0.8 to 1.
On the other hand, I($f^2$) gradually increases with Co concentration, reflecting the gradual increase of hybridization strength as in the analysis results of Ce 3$d_{5/2}$ core-level spectra (Fig. 2(b)).

\begin{figure}
\includegraphics[width=75mm,clip]{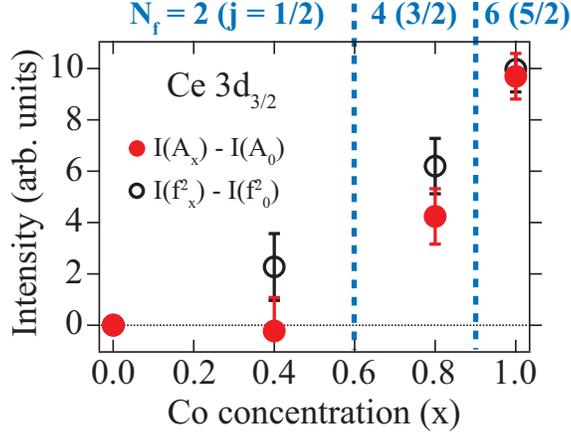}
\caption{\label{fig:figure4} (color online)
Differences of the A peaks (solid circle) and those of the $f^2$ peaks (open circle) obtained from the differences of Ce 3$d_{3/2}$ core-level spectra between CeNiGe$_2$ ($x$ = 0) and the other samples ($x$ = 0.4, 0.8, 1). Each value exhibits the different tendencies; while $f^2$ peaks gradually change, A peaks suddenly change between $x$ = 0.4 and 0.8. The degeneracy (N$_f$) and total angular momentum of $f$-electrons (j) are depicted above the panel}
\end{figure}

Figure 4 is the plot of intensity difference (I(A$_x$)-I(A$_0$) and I($f^0_x$)-I($f^0_0$), where the subscript represents Co concentration, $x$).
The degeneracy in ground state varies due to CEF effects as indicated above the panel.
It is recognized that the variation of A peak (solid circle) accompanies that of the degeneracy:
note that N$_f$ = 2 for $x$ = 0 and 0.4, N$_f$ = 4 for $x$ = 0.8, and N$_f$ = 6 for $x$ =1.
Moreover, measurement temperature (T = 20 K) is enough low to consider that the above behaviors reflect the properties of the ground state;
the energy difference between the ground state and first excited state due to CEF effects is about 100 K for $x$ = 0, 0.4 and is about 60 K for $x$ = 0.8 from the specific heat measurements \cite{Lee05}.
These results certainly indicate that the variation of multiplet structures in Ce 3$d_{3/2}$ $f^1$ peak is attributed to the variation of degeneracy in the ground state due to CEF effects.
Here, we should carefully discuss that the variation of multiplet structure solely originates from the change of degeneracy, because the change of degeneracy also influence on hybridization strength which can affect the multiplet structures.
Firstly, it should be noted that our data reveal the different tendency between A peak and $f^2$ peak in Ce 3$d_{3/2}$ spectra as shown in Fig. 3 and 4.
This indicates that the origin of variation of A peak is not the same to that of $f^2$ which reflect the variation of hybridization strength.
The change of A peak is accompanied with that of degeneracy as mentioned above.  
Next, let us compare the obtained HXPES results with RPES and theoretical studies.
In Fig. 2(b), both hybridization strength from HXPES and RPES increase relatively much from $x$ = 0.6 to 0.8 comparing to the case of from $x$ = 0 to 0.6.
However, the quantity of variation is $\sim$ 1 meV in the results of RPES (16.5 for $x$ = 0.6 to 17.5 meV for $x$ = 0.8).
This value seems to be too small to affect the change of multiplet structure.
In fact, the large change of Kondo temperature is also the result of consideration of degeneracy as in the results of RPES \cite{Bick87,Im05}.
In addition, theoretical studies, which are based on the SIAM considering atomic multiples, report that the change of I(A)/I(B) in multiplet structure are much considerable in the case of the change of degeneracy rather than in the case of the hybridization strength \cite{Imada89}.
From the above discussions, we conclude that the large change of the multiplet structure in Ce 3$d_{3/2}$ $f^1$ peak mainly comes from the change of degeneracy rather than that of hybridization strength.

\section{summary}

In summary, the bulk Ce 3$d$ core-level spectra of CeNi$_{1-x}$Co$_x$Ge$_2$ (0 $\leq$ $x$ $\leq$ 1) were systematically investigated to clarify the correlation effects between electrons due to CEF effects, using high-resolution hard X-ray photoemission spectroscopy.
Satellite structures of Ce 3$d_{5/2}$ core-level spectra well explain the change of the hybridization strength in good agreement with the results of Ce 3$d$-4$f$ and 4$d$-4$f$ RPES.
The variation of multiplet structures in the $f^1$ peak of Ce 3$d_{3/2}$ core-level spectra dominantly originates from the change of the degeneracy of $f$-state in the ground state due to CEF effects.
These results indicate that the multiplet structure is essential to study the strong correlation effects between electrons, e.g., CEF effects, in heavy-fermion Ce compounds.


\begin{acknowledgments}
We acknowledge the support of J. J. Kim and E. Ikenaga at SPring-8 and the helpful discussion of H. Ogasawara at Tohoku University. 
This work is funded by the Korean Goverment (KRF-2008-313-C00293) and by Grant-in-Aid of Scientific Research (B) (No.18340110) from MEXT of Japan, and is performed for the Nuclear R\&D Programs funded by the Ministry of Science
\& Technology of Korea.
\end{acknowledgments}

\bibliography{reference}

\begin{thebibliography}{19}
\expandafter\ifx\csname natexlab\endcsname\relax\def\natexlab#1{#1}\fi
\expandafter\ifx\csname bibnamefont\endcsname\relax
  \def\bibnamefont#1{#1}\fi
\expandafter\ifx\csname bibfnamefont\endcsname\relax
  \def\bibfnamefont#1{#1}\fi
\expandafter\ifx\csname citenamefont\endcsname\relax
  \def\citenamefont#1{#1}\fi
\expandafter\ifx\csname url\endcsname\relax
  \def\url#1{\texttt{#1}}\fi
\expandafter\ifx\csname urlprefix\endcsname\relax\def\urlprefix{URL }\fi
\providecommand{\bibinfo}[2]{#2}
\providecommand{\eprint}[2][]{\url{#2}}

\bibitem[{\citenamefont{Hufner}(1995)}]{Hufn95}
\bibinfo{author}{\bibfnamefont{S.}~\bibnamefont{Hufner}},
  \emph{\bibinfo{title}{Photoelectron Spectroscopy}}
  (\bibinfo{publisher}{Springer-Verlag}, \bibinfo{address}{Berlin},
  \bibinfo{year}{1995}).

\bibitem[{\citenamefont{Sakai et~al.}(1988)\citenamefont{Sakai, Motizuki, and
  Kasuya}}]{Saka88}
\bibinfo{author}{\bibfnamefont{O.}~\bibnamefont{Sakai}},
  \bibinfo{author}{\bibfnamefont{M.}~\bibnamefont{Motizuki}}, \bibnamefont{and}
  \bibinfo{author}{\bibfnamefont{T.}~\bibnamefont{Kasuya}},
  \emph{\bibinfo{title}{Core-Level Spectroscopy in Condensed Systems}}
  (\bibinfo{publisher}{Springer-Verlag}, \bibinfo{address}{Berlin Heidelberg},
  \bibinfo{year}{1988}).

\bibitem[{\citenamefont{Gunnarsson and Schonhammer}(1983)}]{Gunn83}
\bibinfo{author}{\bibfnamefont{O.}~\bibnamefont{Gunnarsson}} \bibnamefont{and}
  \bibinfo{author}{\bibfnamefont{K.}~\bibnamefont{Schonhammer}},
  \bibinfo{journal}{Phys. Rev. B} \textbf{\bibinfo{volume}{28}},
  \bibinfo{pages}{4315} (\bibinfo{year}{1983}).

\bibitem[{\citenamefont{Kim et~al.}(2004)\citenamefont{Kim, Noh, Kim, and
  Oh}}]{Kim04}
\bibinfo{author}{\bibfnamefont{H.-D.} \bibnamefont{Kim}},
  \bibinfo{author}{\bibfnamefont{H.-J.} \bibnamefont{Noh}},
  \bibinfo{author}{\bibfnamefont{K.~H.} \bibnamefont{Kim}}, \bibnamefont{and}
  \bibinfo{author}{\bibfnamefont{S.-J.} \bibnamefont{Oh}},
  \bibinfo{journal}{Phys. Rev. Lett.} \textbf{\bibinfo{volume}{93}},
  \bibinfo{pages}{126404} (\bibinfo{year}{2004}).

\bibitem[{\citenamefont{Taguchi et~al.}(2005)\citenamefont{Taguchi, Chainani,
  Horiba, Takata, Yabashi, Tamasaku, Nishino, Miwa, Ishikawa, Takeuchi
  et~al.}}]{Tagu05}
\bibinfo{author}{\bibfnamefont{M.}~\bibnamefont{Taguchi}},
  \bibinfo{author}{\bibfnamefont{A.}~\bibnamefont{Chainani}},
  \bibinfo{author}{\bibfnamefont{K.}~\bibnamefont{Horiba}},
  \bibinfo{author}{\bibfnamefont{Y.}~\bibnamefont{Takata}},
  \bibinfo{author}{\bibfnamefont{M.}~\bibnamefont{Yabashi}},
  \bibinfo{author}{\bibfnamefont{K.}~\bibnamefont{Tamasaku}},
  \bibinfo{author}{\bibfnamefont{Y.}~\bibnamefont{Nishino}},
  \bibinfo{author}{\bibfnamefont{D.}~\bibnamefont{Miwa}},
  \bibinfo{author}{\bibfnamefont{T.}~\bibnamefont{Ishikawa}},
  \bibinfo{author}{\bibfnamefont{T.}~\bibnamefont{Takeuchi}},
  \bibnamefont{et~al.}, \bibinfo{journal}{Phys. Rev. Lett.}
  \textbf{\bibinfo{volume}{95}}, \bibinfo{pages}{177002}
  (\bibinfo{year}{2005}).

\bibitem[{\citenamefont{Cox and Zawadowski}(1999)}]{Cox99}
\bibinfo{author}{\bibfnamefont{D.~L.} \bibnamefont{Cox}} \bibnamefont{and}
  \bibinfo{author}{\bibfnamefont{A.}~\bibnamefont{Zawadowski}},
  \emph{\bibinfo{title}{Exotic Kondo effects in metals: Magnetic ions in a
  crystalline electric field and tunnelling centres}}
  (\bibinfo{publisher}{Taylor and Francis}, \bibinfo{year}{1999}).

\bibitem[{\citenamefont{Takimoto et~al.}(2002)\citenamefont{Takimoto, Hotta,
  Maehira, and Ueda}}]{Taki02}
\bibinfo{author}{\bibfnamefont{T.}~\bibnamefont{Takimoto}},
  \bibinfo{author}{\bibfnamefont{T.}~\bibnamefont{Hotta}},
  \bibinfo{author}{\bibfnamefont{T.}~\bibnamefont{Maehira}}, \bibnamefont{and}
  \bibinfo{author}{\bibfnamefont{K.}~\bibnamefont{Ueda}}, \bibinfo{journal}{J.
  Phys.: Condens. Matter} \textbf{\bibinfo{volume}{14}}, \bibinfo{pages}{L369}
  (\bibinfo{year}{2002}).

\bibitem[{\citenamefont{Curro et~al.}(2001)\citenamefont{Curro, Simovic,
  Hammel, Pagliuso, Sarrao, Thompson, and Martins}}]{Curr01}
\bibinfo{author}{\bibfnamefont{N.~J.} \bibnamefont{Curro}},
  \bibinfo{author}{\bibfnamefont{B.}~\bibnamefont{Simovic}},
  \bibinfo{author}{\bibfnamefont{P.~C.} \bibnamefont{Hammel}},
  \bibinfo{author}{\bibfnamefont{P.~G.} \bibnamefont{Pagliuso}},
  \bibinfo{author}{\bibfnamefont{J.~L.} \bibnamefont{Sarrao}},
  \bibinfo{author}{\bibfnamefont{J.~D.} \bibnamefont{Thompson}},
  \bibnamefont{and} \bibinfo{author}{\bibfnamefont{G.~B.}
  \bibnamefont{Martins}}, \bibinfo{journal}{Phys. Rev. B}
  \textbf{\bibinfo{volume}{64}}, \bibinfo{pages}{180514(R)}
  (\bibinfo{year}{2001}).

\bibitem[{\citenamefont{Lee et~al.}(2005)\citenamefont{Lee, Hong, Kim, Jang,
  Mun, Jung, Kimura, Park, Park, and Kwon}}]{Lee05}
\bibinfo{author}{\bibfnamefont{B.~K.} \bibnamefont{Lee}},
  \bibinfo{author}{\bibfnamefont{J.~B.} \bibnamefont{Hong}},
  \bibinfo{author}{\bibfnamefont{J.~W.} \bibnamefont{Kim}},
  \bibinfo{author}{\bibfnamefont{K.-H.} \bibnamefont{Jang}},
  \bibinfo{author}{\bibfnamefont{E.~D.} \bibnamefont{Mun}},
  \bibinfo{author}{\bibfnamefont{M.~H.} \bibnamefont{Jung}},
  \bibinfo{author}{\bibfnamefont{S.}~\bibnamefont{Kimura}},
  \bibinfo{author}{\bibfnamefont{T.}~\bibnamefont{Park}},
  \bibinfo{author}{\bibfnamefont{J.-G.} \bibnamefont{Park}}, \bibnamefont{and}
  \bibinfo{author}{\bibfnamefont{Y.~S.} \bibnamefont{Kwon}},
  \bibinfo{journal}{Phys. Rev. B} \textbf{\bibinfo{volume}{71}},
  \bibinfo{pages}{214433} (\bibinfo{year}{2005}).

\bibitem[{\citenamefont{Doniach}(1977)}]{Doni77}
\bibinfo{author}{\bibfnamefont{S.}~\bibnamefont{Doniach}},
  \bibinfo{journal}{Physica B} \textbf{\bibinfo{volume}{91B}},
  \bibinfo{pages}{231} (\bibinfo{year}{1977}).

\bibitem[{\citenamefont{Im et~al.}(2006)\citenamefont{Im, Ito, Kimura, Hong,
  and Kwon}}]{Im06}
\bibinfo{author}{\bibfnamefont{H.~J.} \bibnamefont{Im}},
  \bibinfo{author}{\bibfnamefont{T.}~\bibnamefont{Ito}},
  \bibinfo{author}{\bibfnamefont{S.}~\bibnamefont{Kimura}},
  \bibinfo{author}{\bibfnamefont{J.~B.} \bibnamefont{Hong}}, \bibnamefont{and}
  \bibinfo{author}{\bibfnamefont{Y.~S.} \bibnamefont{Kwon}},
  \bibinfo{journal}{Physica B} \textbf{\bibinfo{volume}{378-380}},
  \bibinfo{pages}{825} (\bibinfo{year}{2006}).

\bibitem[{\citenamefont{Tanuma et~al.}(1987)\citenamefont{Tanuma, Powell, and
  Penn}}]{Tanu87}
\bibinfo{author}{\bibfnamefont{S.}~\bibnamefont{Tanuma}},
  \bibinfo{author}{\bibfnamefont{C.~J.} \bibnamefont{Powell}},
  \bibnamefont{and} \bibinfo{author}{\bibfnamefont{D.~R.} \bibnamefont{Penn}},
  \bibinfo{journal}{Surf. Sci.} \textbf{\bibinfo{volume}{192}},
  \bibinfo{pages}{L847} (\bibinfo{year}{1987}).

\bibitem[{\citenamefont{Shirley}(1972)}]{Shir72}
\bibinfo{author}{\bibfnamefont{D.~A.} \bibnamefont{Shirley}},
  \bibinfo{journal}{Phys. Rev. B} \textbf{\bibinfo{volume}{5}},
  \bibinfo{pages}{4709} (\bibinfo{year}{1972}).

\bibitem[{\citenamefont{Fuggle et~al.}(1983)\citenamefont{Fuggle, Hillebrecht,
  Zolnierek, Lasser, Freiburg, Gunnarsson, and Schonhammer}}]{Fugg83}
\bibinfo{author}{\bibfnamefont{J.~C.} \bibnamefont{Fuggle}},
  \bibinfo{author}{\bibfnamefont{F.~U.} \bibnamefont{Hillebrecht}},
  \bibinfo{author}{\bibfnamefont{Z.}~\bibnamefont{Zolnierek}},
  \bibinfo{author}{\bibfnamefont{R.}~\bibnamefont{Lasser}},
  \bibinfo{author}{\bibfnamefont{C.}~\bibnamefont{Freiburg}},
  \bibinfo{author}{\bibfnamefont{O.}~\bibnamefont{Gunnarsson}},
  \bibnamefont{and}
  \bibinfo{author}{\bibfnamefont{K.}~\bibnamefont{Schonhammer}},
  \bibinfo{journal}{Phys. Rev. B} \textbf{\bibinfo{volume}{27}},
  \bibinfo{pages}{7330} (\bibinfo{year}{1983}).

\bibitem[{\citenamefont{Braicovich et~al.}(1997)\citenamefont{Braicovich,
  Brookes, Dallera, Salvietti, and Olcese}}]{Brai97}
\bibinfo{author}{\bibfnamefont{L.}~\bibnamefont{Braicovich}},
  \bibinfo{author}{\bibfnamefont{N.~B.} \bibnamefont{Brookes}},
  \bibinfo{author}{\bibfnamefont{C.}~\bibnamefont{Dallera}},
  \bibinfo{author}{\bibfnamefont{M.}~\bibnamefont{Salvietti}},
  \bibnamefont{and} \bibinfo{author}{\bibfnamefont{G.~L.}
  \bibnamefont{Olcese}}, \bibinfo{journal}{Phys. Rev. B}
  \textbf{\bibinfo{volume}{56}}, \bibinfo{pages}{15047} (\bibinfo{year}{1997}).

\bibitem[{\citenamefont{Imada and Jo}(1989)}]{Imada89}
\bibinfo{author}{\bibfnamefont{S.}~\bibnamefont{Imada}} \bibnamefont{and}
  \bibinfo{author}{\bibfnamefont{T.}~\bibnamefont{Jo}}, \bibinfo{journal}{Phys.
  Soc. Jpn.} \textbf{\bibinfo{volume}{58}}, \bibinfo{pages}{2665}
  (\bibinfo{year}{1989}).

\bibitem[{\citenamefont{Matsunami et~al.}(2008)\citenamefont{Matsunami, Horiba,
  Taguchi, Yamamoto, Chainani, Takata, Senba, Ohashi, Yabashi, Tamasaku
  et~al.}}]{Mats08}
\bibinfo{author}{\bibfnamefont{M.}~\bibnamefont{Matsunami}},
  \bibinfo{author}{\bibfnamefont{K.}~\bibnamefont{Horiba}},
  \bibinfo{author}{\bibfnamefont{M.}~\bibnamefont{Taguchi}},
  \bibinfo{author}{\bibfnamefont{K.}~\bibnamefont{Yamamoto}},
  \bibinfo{author}{\bibfnamefont{A.}~\bibnamefont{Chainani}},
  \bibinfo{author}{\bibfnamefont{Y.}~\bibnamefont{Takata}},
  \bibinfo{author}{\bibfnamefont{Y.}~\bibnamefont{Senba}},
  \bibinfo{author}{\bibfnamefont{H.}~\bibnamefont{Ohashi}},
  \bibinfo{author}{\bibfnamefont{M.}~\bibnamefont{Yabashi}},
  \bibinfo{author}{\bibfnamefont{K.}~\bibnamefont{Tamasaku}},
  \bibnamefont{et~al.}, \bibinfo{journal}{Phys. Rev. B}
  \textbf{\bibinfo{volume}{77}}, \bibinfo{pages}{165126}
  (\bibinfo{year}{2008}).

\bibitem[{\citenamefont{Im et~al.}(2005)\citenamefont{Im, Ito, Hong, Kimura,
  and Kwon}}]{Im05}
\bibinfo{author}{\bibfnamefont{H.~J.} \bibnamefont{Im}},
  \bibinfo{author}{\bibfnamefont{T.}~\bibnamefont{Ito}},
  \bibinfo{author}{\bibfnamefont{J.~B.} \bibnamefont{Hong}},
  \bibinfo{author}{\bibfnamefont{S.~I.} \bibnamefont{Kimura}},
  \bibnamefont{and} \bibinfo{author}{\bibfnamefont{Y.~S.} \bibnamefont{Kwon}},
  \bibinfo{journal}{Phys. Rev. B} \textbf{\bibinfo{volume}{72}},
  \bibinfo{pages}{220405(R)} (\bibinfo{year}{2005}).

\bibitem[{\citenamefont{Bickers et~al.}(1987)\citenamefont{Bickers, Cox, and
  Wilkins}}]{Bick87}
\bibinfo{author}{\bibfnamefont{N.~E.} \bibnamefont{Bickers}},
  \bibinfo{author}{\bibfnamefont{D.~L.} \bibnamefont{Cox}}, \bibnamefont{and}
  \bibinfo{author}{\bibfnamefont{J.~W.} \bibnamefont{Wilkins}},
  \bibinfo{journal}{Phys. Rev. B} \textbf{\bibinfo{volume}{36}},
  \bibinfo{pages}{2036} (\bibinfo{year}{1987}).

\end{thebibliography}

\end{document}